\documentclass[12pt,aps,eqsecnum,nofootinbib,floatfix]{revtex4}
\usepackage{pifont}
\usepackage{subfigure}
\usepackage{epsfig}
\usepackage{CJK}
\usepackage{amsmath}
\usepackage{amsfonts}
\usepackage{amssymb}
\usepackage{color}
\usepackage{graphicx}
\usepackage{mathrsfs}

\usepackage[normalem]{ulem}
\def\be{\begin{equation}}
\def\ee{\end{equation}}
\def\ba{\begin{eqnarray}}
\def\ea{\end{eqnarray}}

\definecolor{dyellow}{rgb}{1.,0.8,.0}
\definecolor{myblue}{rgb}{.1,.1,.7}
\definecolor{dcyan}{rgb}{.0,.6,.6}
\definecolor{dmagenta}{rgb}{0.6,0.0,0.6}
\definecolor{brown}{rgb}{0.6,0.2,0.}
\definecolor{darkblue}{rgb}{.0,.0,0.5}
\definecolor{darkred}{rgb}{0.75,0.0,0.0}
\definecolor{orange}{rgb}{1.,.6,.0}
\definecolor{dorange}{rgb}{0.8,.4,.0}
\definecolor{darkgreen}{rgb}{0.0,0.6,0.0}
\definecolor{purple}{rgb}{.4,.0,.4}
\definecolor{lightgrey}{rgb}{0.7, 0.7, 0.7}
\definecolor{grey}{rgb}{0.4, 0.4, 0.4}
\newcommand{\bea}{\begin{eqnarray}}
\newcommand{\eea}{\end{eqnarray}}

\newcommand\btd{\raise 2pt
\hbox{$\hat\bigtriangledown$}\hskip 1.5pt}
\newcommand\bt{\raise 2pt
\hbox{$\bigtriangledown$}\hskip 1.5pt}

\newcommand{\omits}[1]{}

\def\PRD{{Phys. Rev. D}}

\def\PLB{{Phys. Lett.}~{\bf B}}

\def\CQG{{Class. Quant. Grav. }}

\begin{document}%\large\bf
\begin{CJK*}{GBK}{song}

\title{Revisit on the thermodynamic stability of Ho\v{r}ava-Lifshitz black hole}

\author{Xudong Meng$^{a,b}$
\footnote{The corresponding author, Email: mengxd2000@126.com},
Ruihong Wang$^c$\footnote{Email: ruihong\_wang@126.com}}

\medskip

\affiliation{ \footnotesize$^a$College of Science of Hebei North University, 075000 Hebei, China\\
\footnotesize$^b$Institute of New Energy Science and Technology of Hebei North Univesity, 075000 Hebei,
China\\
\footnotesize$^c$College of Information Science and Technology, Hebei Agricultural University, Baoding 071001, China}

\begin{abstract}

We study the thermodynamic properties of the black hole derived in Ho\v{r}ava-Lifshitz (HL) gravity without the detailed-balance condition.
%This black hole has the similar structure to that of the RN-AdS black hole.
The parameter $\Xi=\epsilon^2$ in the HL black hole plays the same role as that of the electric charge in the RN-AdS black hole.
By analogy, we treat the parameter $\Xi$ as the thermodynamic variable and obtain the first law of thermodynamics for the HL black hole. Although the HL black hole and the RN-AdS black hole have the similar mass and temperature, due to their very different entropy the two black holes have very different thermodynamic properties. 
By calculating the heat capacity and the free energy we analyze the thermodynamic stability of the HL black hole.

\end{abstract}

%\pacs{04.50.-h, 04.20.Jb, 04.90.+e}

\maketitle

%\tableofcontents
\bigskip

%%%%%%%%%%%%%%%%%%%%%%%%%%%%%%%%%%%%%%%%%%%%%%%%%%%%%%%%%%%%%%%%%%%%%%%%
%%%%%%%%%%%%%%%%%%%%%%%%%%%%%%%%%%%%%%%%%%%%%%%%%%%%%%%%%%%%%%%%%%%%%%%%
%%%%%%%%%                                            %%%%%%%%%%%%%%%%%%%
%%%%%%%%%               I.  A Review                 %%%%%%%%%%%%%%%%%%%
%%%%%%%%%                                            %%%%%%%%%%%%%%%%%%%
%%%%%%%%%%%%%%%%%%%%%%%%%%%%%%%%%%%%%%%%%%%%%%%%%%%%%%%%%%%%%%%%%%%%%%%%
%%%%%%%%%%%%%%%%%%%%%%%%%%%%%%%%%%%%%%%%%%%%%%%%%%%%%%%%%%%%%%%%%%%%%%%%

\section{Introduction}

%The discovery of Hawking radiation of black holes arouses much interest in the black hole thermodynamics\cite{Hawking-1975}.
Black hole is much like a thermodynamic system. Besides temperature, it has also entropy\cite{Hawking-1975,Bekenstein-1973}. To describe a macroscopic black hole, some other parameters such as black hole mass $M$, electric charge $Q$ and the angular momentum $J$ can be introduced. For black holes in anti-de Sitter space, one can even treat the cosmological constant as the thermodynamic pressure and its conjugated quantity as the thermodynamic volume to construct the extended phase space\cite{Kastor,Dolan,Mann1}. These thermodynamic quantities of black hole satisfy the first law of thermodynamics. Later, it was found that black holes have many other thermodynamic properties, such as phase transition and critical phenomena\cite{Chamblin.1999,Chamblin.1999b,Cvetic.1999b,Peca.1999,Wu.2000,Biswas.2004,Myung.2005,Dey.2007,Myung.2008,Sahay.2010,Banerjee.2011,LiuYX,Ma.2014,Ma.2014b,Ma.2014c}. It is also found that black holes should have the second-order phase transition at the point where the heat capacity diverges by analogy with the conventional thermodynamic system\cite{Davies.1978}. Since the discovery of Hawking-Page phase transition, the critical behaviors of black holes in the asymptotically anti-de Sitter have been extensively studied. Specifically, after introducing the pressure and thermodynamic volume, in the extended phase space it is found that the $P-V$ criticality of the black holes are very similar to that of the van der Waals (VdW) system\cite{Dolan,Mann1,Cai.2013,Chen.2013,Hendi.2013,Altamirano-2014,WangB,Xu.2014,Ma.2015,Ma.2015b,ZhaoHH.2015,Dykaar.2017}.

Ho\v{r}ava-Lifshitz gravity is a kind of power-counting renormalizable gravity theory\cite{Horava.2009}. The exact solutions and the cosmological applications of the HL gravity have been extensively studied. In this theory, the first static, spherically symmetric black hole solution was derived in \cite{Lu.2009}, where the authors also extended the Ho\v{r}ava-Lifshitz (HL) black hole to the case without detailed-balance condition. Considering the different topological structures of the event horizons, more general HL black hole solution is derived in \cite{Cai.2009c,Cai.2009b}. Myung studied the thermodynamics of the Kehagias-Sfetsos black hole\cite{Myung.2009}, which is the solution of the deformed HL gravity. The phase structure and critical behaviors of the charged topological HL black hole were also studied in \cite{Cao.2011,Majhi.2012,Mo.2013}. Using the horizon thermodynamics\cite{Padmanabhan.2002,Ma2}, Ma et.al found a universal phase structure and the $P-V$ criticality for the topological HL black hole\cite{Ma-CQG.2017}. In particular, it is found that for the uncharged HL black hole with the spherical horizon there is a peculiar $P-V$ criticality, which exhibits a critical curve, but not a critical point\cite{Ma-PRD.2017}.

On the basis of the work of Ma. et al, \cite{Ma-PRD.2017},
in this paper we further study the uncharged HL black hole with only the spherical horizon. The parameter $\epsilon^2$ in the HL black hole plays the same role as that of the electric charge $Q$ in the RN-AdS black hole.
We treat the $\Xi=\epsilon^2$ as a new thermodynamic variable and obtain the first law of thermodynamics. In this manner, we can study the thermodynamic stability of the HL black hole.

The paper is arranged as follows: in the next section we simply
introduce the HL black hole solution and the
corresponding thermodynamic quantities. In section 3  we will study the thermodynamic stability by
calculating the heat capacity  and the free energy.
We will make some
concluding remarks in section 4.

\section{Ho\v{r}ava-Lifshitz black hole}

We consider the static, spherically symmetric spacetime, with the metric ansatz,
\be\label{staticmetric}
ds^2 =-\tilde N^2(r)f(r) dt^2 +\frac{dr^2}{f(r)} +r^2
d\Omega^2.
\ee
Here we are only concerned with the black hole solution with spherical horizon.
In this metric ansatz, the Ho\v{r}ava-Lifshitz black hole was derived in the case of no detailed-balance condition \cite{Lu.2009,Cai.2009c}.
It is shown that the metric function $\tilde N(r)=N_0$, which could be set to one,  and  $f(r)$ is given by
\be\label{metric}
f(r)=1+\frac{x^2}{1-\epsilon ^2}-\frac{\sqrt{\left(1-\epsilon ^2\right)m x+x^4 \epsilon ^2}}{1-\epsilon ^2},
\ee
where $x=\sqrt{-\Lambda}r$, is a dimensionless variable. $m$ is an integration constant related to the black hole mass,
\be\label{mq}
M=\frac{\kappa ^2 \mu ^2 \Omega \sqrt{-\Lambda }}{16}  m=\alpha m.
\ee
Here $\Omega$ is the volume of the two-dimensional Einstein space.  We set $\frac{\kappa ^2 \mu ^2 \Omega \sqrt{-\Lambda }}{16}=\alpha$ for short. Obviously, $\Lambda$ must be negative and $\alpha$ must be positive.

%Because only for the case with $\lambda=1$, general relativity can be recovered in the large distance approximation, we only consider $\lambda=1$ in the following.
%In this case, from Eq.(\ref{constants}), one can see that $\Lambda$ must be negative.

$\epsilon=0$ corresponds to the so-called detailed-balance condition. %under which HL gravity turns out to
%be intimately related to topological gravity in three dimensions and the geometry of the Cotton tensor.
In this case, the solution reduces to $f(r)=1+x^2-\sqrt{mx}$, the thermodynamic properties has been studied in detail\cite{Majhi.2012}. For the case with $\epsilon=1$, HL gravity returns back to general relativity and the HL black hole becomes a Schwarzschild-(A)dS black hole. Therefore, as the authors did in \cite{Ma-PRD.2017}, we will consider the values of $\epsilon$ in the region $0\leq \epsilon^2 \leq 1$ below. For simplicity, we set $\Xi=\epsilon^2$ below, thus $0\leq \Xi \leq 1$.

 From Eqs.(\ref{metric}) and (\ref{mq}), the black hole mass is
\be\label{HLM}
M=\frac{\alpha  \left(x_+^4+2 x_+^2+1-\Xi\right)}{x_+},
\ee
where $x_{+}$ is the position of the event horizon of the HL black hole.

According to the metric function, one can easily derive the temperature:
\be\label{tem}
T=\frac{\sqrt{-\Lambda } \left(\Xi +3 x_+^4+2 x_+^2-1\right)}{8 \pi  x_+ \left(x_+^2-\Xi+1\right)}.
\ee
The entropy of the charged topological HL black hole is
\bea\label{entropy}
S=\frac{4 \pi  \alpha }{\sqrt{-\Lambda }} \left[2 \left(1-\Xi \right) \ln \left(x_+\right)+x_+^2\right]+S_0.
\eea
$S_0$ is an integration constant, which cannot be easily determined. For simplicity, we always set $S_0=0$ below. It should be noted that the entropy has the same form even if the electric field exists. In the following, we set $\alpha=1,~ \Lambda=-1$ for simplicity.

To understand the structure of the black hole solution, we expand the metric function for large $r$. It has the form\footnote{Because we set $\Lambda=-1$, in fact $x=\sqrt{-\Lambda} r=r$. But the metric function $f(x)$ does not contain $\Lambda$ explicitly. We still use $x$ instead of $r$ in the following.}
\be\label{expansion}
f(x)=1-\frac{M}{2 \sqrt{\Xi } x}-\frac{M^2 (\Xi -1)}{8 \Xi ^{3/2} x^4}+\frac{x^2}{\sqrt{\Xi }+1}+O(\frac{1}{x^5}).
\ee
For very large $r$, it approaches to
\be
f(x)=1+\frac{x^2}{\sqrt{\Xi }+1},
\ee
which means an asymptotical anti-de Sitter structure. From Eq.(\ref{expansion}), it is found that, except for the $1/x^4$-term, asymptotically this black hole also has a similar structure to that of the RN-AdS black hole, which has the metric function
\be\label{RNAdS}
f(r)=1-\frac{2M}{r}+\frac{Q^2}{r^2}+\frac{r^2}{l^2}.
\ee
This ensures that the HL black hole has the similar thermodynamic properties to that of RN-AdS black hole, such as $P-V$ criticality\cite{Ma-PRD.2017}. In Eq.(\ref{expansion}) there are only two parameters $(M, ~\Xi)$, whereas in Eq.(\ref{RNAdS}) there are three parameters $(M, ~Q,~l)$. It is just this feature that leads to the occurrence of the ``critical curve" in the $P-V$ criticality of the HL black hole\cite{Ma-PRD.2017}. $\Xi$  plays both the roles of $(Q,~l)$.

For further comparison, we list other thermodynamic quantities of the RN-AdS black hole,
\be\label{RNM}
M=\frac{1}{2}\left(r_{+}+\frac{Q^2}{r_{+}}+\frac{r_{+}^3}{l^2}\right), \quad T=\frac{-l^2 Q^2+l^2 r_+^2+3 r_+^4}{4 \pi  l^2 r_+^3}, \quad S=\pi r_{+}^2.
\ee
Clearly, the mass of the RN-AdS black hole is very similar to the mass of the HL black hole, Eq.(\ref{HLM}). As for the temperature, the denominators of the two expressions are slightly different. However, the entropy has completely different form for the two black holes. Thus, we anticipate that for the HL black hole there must be some thermodynamic properties different from those of the RN-AdS black hole.

For the RN-AdS black hole, the first law of black hole thermodynamics is
\be
dM=TdS+\Phi dQ,
\ee
with $\Phi=Q/r_{+}$ the electric potential measured at infinity with respect to the event horizon. One can also treat the cosmological as the pressure and construct the extended phase space. But in this paper, we are only concerned with the non-extended phase space.

Due to the similarity between the HL black hole and the RN-AdS black hole, we try to consider the parameter $\Xi$ as another thermodynamic variable. Thus, the first law for the HL black hole is
\be\label{1st}
dM=TdS+\phi d\Xi,
\ee
where $\phi$ is the conjugated quantity to $\Xi$. One can easily obtain $\phi$ according to the first law
\be\label{phi}
\phi=\left.\frac{\partial M}{\partial \Xi}\right|_S=\frac{(\Xi -1) \left(\ln x_++1\right)+3 x_+^4 \ln x_++x_+^2 \left(2 \ln x_+-1\right)}{x_+ \left(-\Xi +x_+^2+1\right)}.
\ee
Obviously, the $\phi$ is much more complicated that the electric potential $\Phi$.

It is also should be noted that there is no Smarr-like relation for the HL black hole due to the logarithmic term in the entropy.

\section{Thermodynamic stability of HL black holes}

\begin{figure}[htp]
\center{
\subfigure[]{
\includegraphics[width=8cm,keepaspectratio]{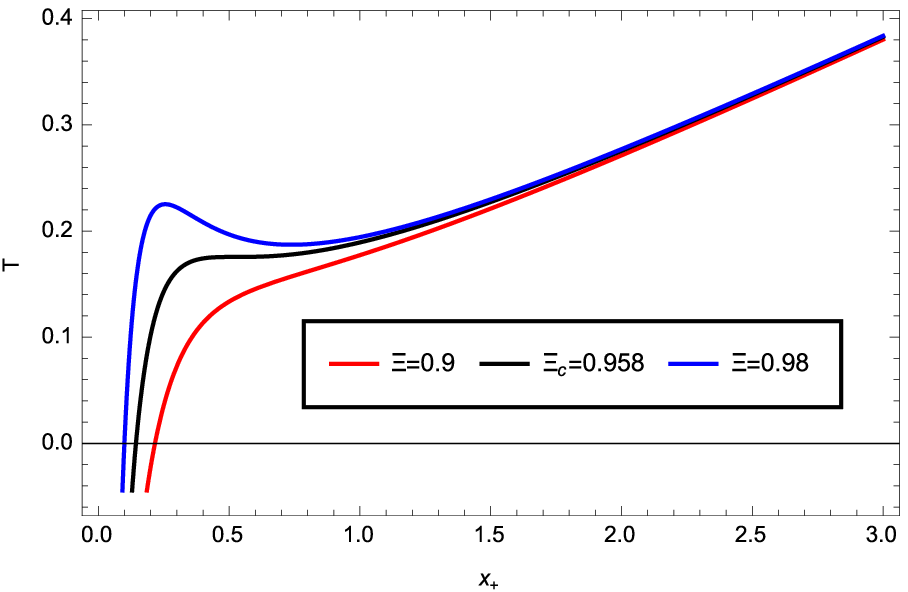}\label{Tx}} \hspace{0.5cm}
\subfigure[]{
\includegraphics[width=8cm,keepaspectratio]{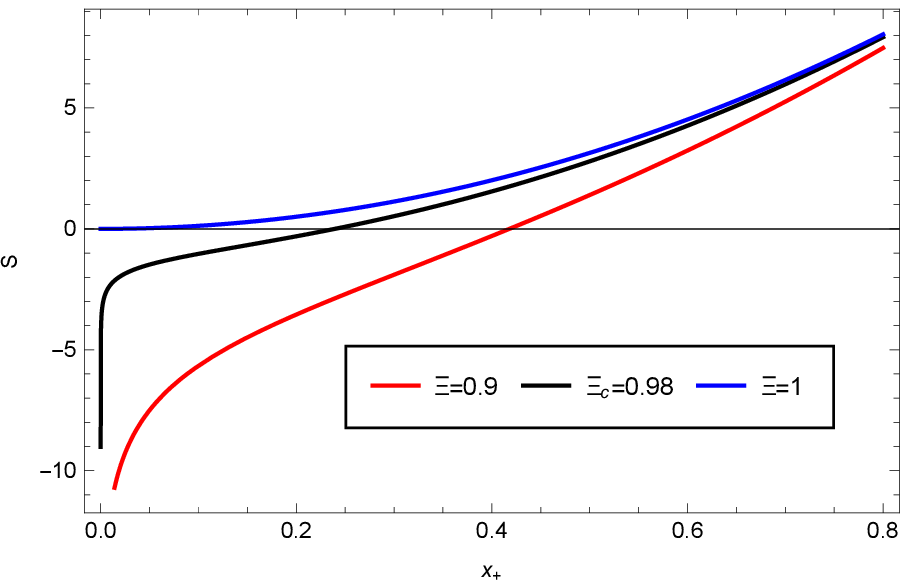}\label{Sx}}
\caption{The temperature and the entropy of the HL black hole as functions of $x_{+}$, respectively. }}
\end{figure}

In this section we will analyze the thermodynamic stability of the HL black hole.
First we analyze the behaviors of the temperature and the entropy of the HL black hole.
As is depicted in Fig.\ref{Tx}, the temperature exhibits a kind of criticality, which is similar to that of RN-AdS black hole.
We can derive the critical point by
\be
\frac{\partial T}{\partial x_{+}}=0, \quad  \frac{\partial^2 T}{\partial x_{+}^2}=0.
\ee
The critical point at $x_{+c}=0.523, ~ \Xi_c=0.958$ and $T_c=0.176$. Here $\Xi$ plays the role of the electric charge $Q$ of the RN-AdS black hole. When $\Xi <\Xi_c$, the temperature is a monotonical function of $x_{+}$. When $\Xi >\Xi_c$, the temperature has a local maximum and a local minimum at two different points. When $\Xi = \Xi_c$, the two extrema coincide at $x_{+}=x_{+c}$.

\begin{figure}[htp]
\center{
\subfigure[]{
\includegraphics[width=8cm,keepaspectratio]{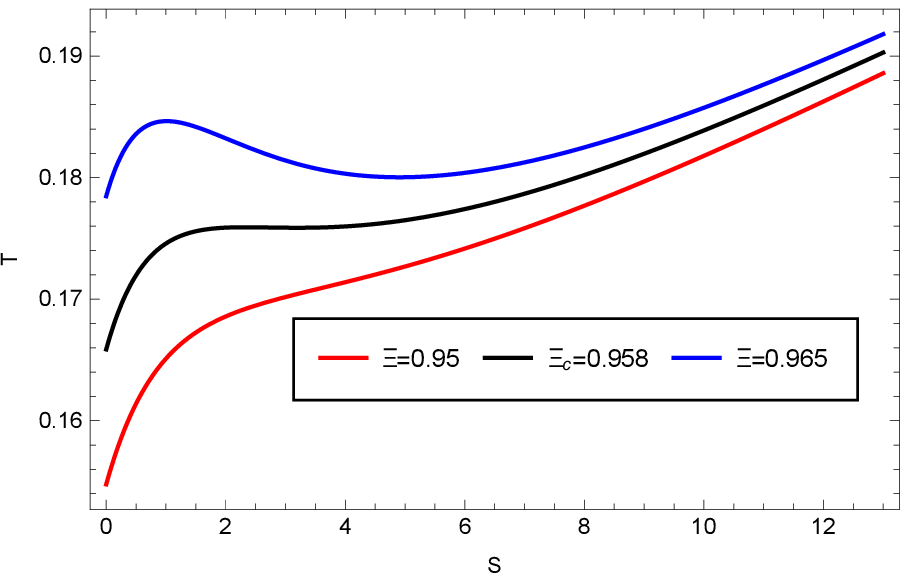}\label{TS}}\hspace{0.5cm}
\subfigure[]{
\includegraphics[width=8cm,keepaspectratio]{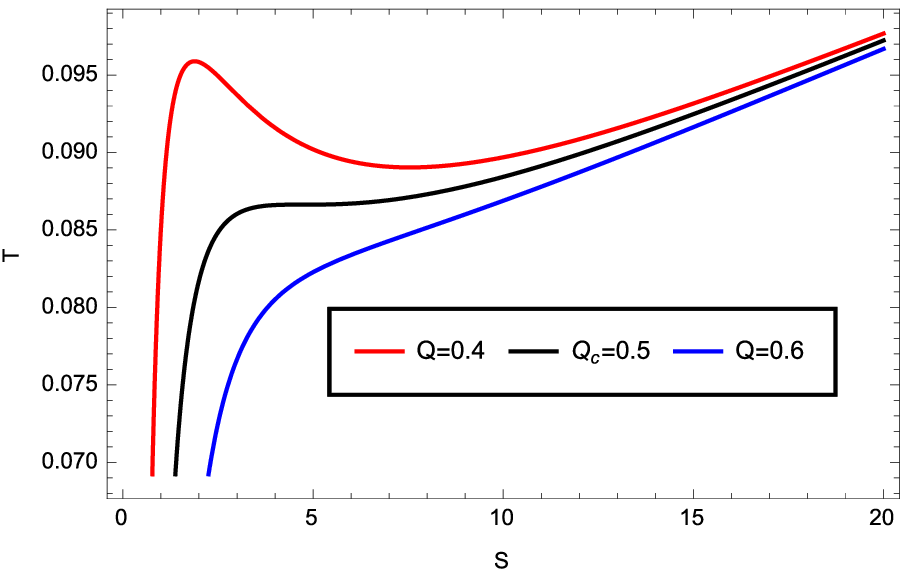}\label{RNTS}}
\caption{The $T-S$ criticality for the HL black hole (the left panel) and the RN-AdS black hole (the right panel). For the RN-AdS black hole, we set $L=3$. }}
\end{figure}

When $0 \leq \Xi < 1$, we find that the entropy of the HL black hole is not always positive. This has been shown in Fig.\ref{Sx}. This property of the HL black hole is not like that of the RN-AdS black hole, for which the entropy is always positive. To be physically meaningful, the temperature and the entropy must be both nonnegative. For clarity, we plot the $T-S$ curves for the HL black hole in Fig.\ref{TS}. It is shown that the temperature has a finite, nonzero value $T_0$ when the entropy tends to zero. The value of $T_0$ is related to the choice of $\Xi$. However, even $\Xi=0$, the $T_0$ is greater than zero when the entropy is zero. By analogy, we also plot the $T-S$ curves for the RN-AdS black hole in Fig.\ref{RNTS}. Clearly, for the RN-AdS black hole the temperature can be zero (the extremal case), while the entropy is always positive.

For this pair of conjugated quantity $(T,~S)$, there is still the criticality for the HL black hole. The critical point is the same as that of the $T-x_{+}$ case above.  As is well known, the VdW liquid/gas system has the $P-V$ criticality, and the pressure $P$ will approach infinity when the volume $V$ tends to zero. Taking $\beta=1/T$, we see that the $\beta-S$ behavior for the RN-AdS black hole is similar to the $P-V$ behaviors of the VdW liquid/gas system. However, the $\beta-S$ curves for the HL black hole exhibit a kind of {\it pseudo}-Van der Waals behavior. By ``pseudo", we mean that for the HL black hole $\beta$ tends to a finite value but not infinity for very small entropy. For larger entropy, the behaviors of the $\beta-S$ curves are the same as that of RN-AdS black hole and the VdW liquid/gas system.

We can calculate the heat capacity to clarify the thermodynamic stability of the HL black hole. According to the first law,
\be
C_{\Xi}=\left.\frac{\partial M}{\partial T}\right|_\Xi=T\left.\frac{\partial S}{\partial T}\right|_\Xi=\frac{8 \pi  \left(-\Xi +x_+^2+1\right){}^2 \left(\Xi +3 x_+^4+2 x_+^2-1\right)}{(\Xi -1)^2+(7-9 \Xi ) x_+^4-5 (\Xi -1) x_+^2+3 x_+^6}.
\ee

\begin{figure}[htp]
\center{
\includegraphics[width=8cm,keepaspectratio]{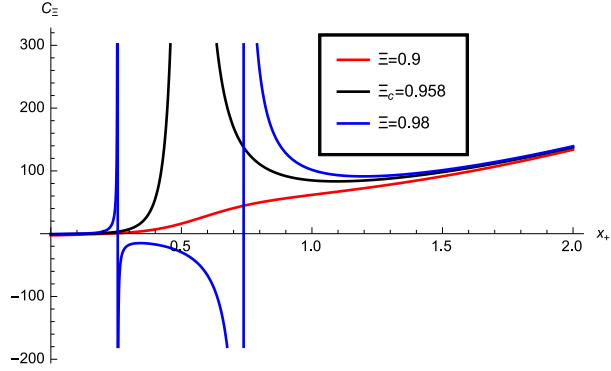}
\caption{The behaviors of the heat capacity of HL black hole. }\label{Cx}}
\end{figure}

As is shown in Fig.\ref{Cx}, the heat capacity is always positive when $\Xi <\Xi_c$. In this case the HL black hole is always thermodynamically stable. When $\Xi >\Xi_c$, the heat capacity will diverge at the two extrema of the temperature. This means that there is the second-order phase transition at the divergent points. In this case, for the large and small black holes, the heat capacities are both positive, whereas, in the intermediate region the heat capacity is negative. The two phases with the positive heat capacity are both locally thermodynamically stable. In fact, for small $x_{+}$ the heat capacity is meaningless due to the negative temperature or negative entropy. However, this will not influence our result given above.

\begin{figure}[htp]
\center{
\includegraphics[width=7cm,keepaspectratio]{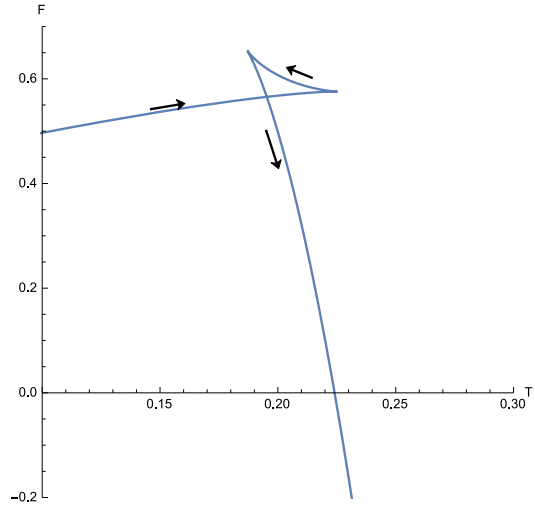}
\caption{The free energy $F$ as function of temperature $T$. Here we take $\Xi=0.98$.}\label{F}}
\end{figure}

To judge the global thermodynamic stability of the HL black hole, we should consult the free energy, which should be defined by
\be
F=M-TS=\frac{-\Xi +x_+^4+2 x_+^2+1}{x_+}-\frac{\left(\Xi +3 x_+^4+2 x_+^2-1\right) \left[2 (1-\Xi ) \ln x_++x_+^2\right]}{2 x_+ \left(-\Xi +x_+^2+1\right)}.
\ee
Fig.\ref{F} shows the behaviors of the free energy $F$.
The arrows point to the direction of increasing $x_{+}$. As we all know, the lower the free energy is, the more stable the thermodynamic system will be.
For fixed temperature , it can be seen that the small black hole has lower free energy and is thermodynamically more stable when the temperature is low. When the temperature is high, the large black hole is more stable.

\begin{figure}[htp]
\center{
\includegraphics[width=8cm,keepaspectratio]{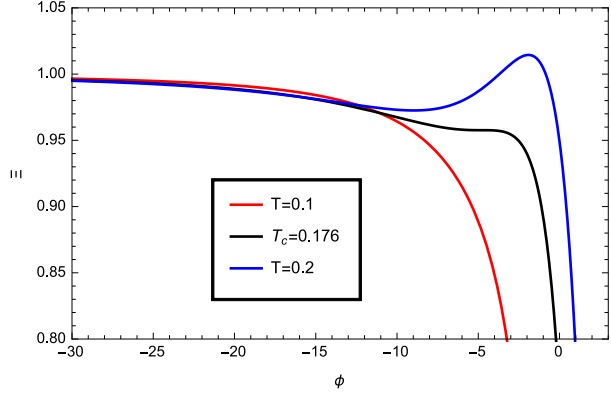}
\caption{The $\Xi-\phi$ isotherms of the HL black hole.}\label{XP}}
\end{figure}

For the RN-AdS black hole, it is known that there is also the $Q-\Phi$ criticality similar to that of the van der Waals system\cite{Wu.2000}. So we try to analyze the relations between $\Xi$ and $\phi$ for fixed temperature.
The equation of state in this case is
\be
\Xi=\frac{\left(1+K-\frac{3}{64 \pi ^2 K T^2}\right) \left(\frac{1}{64 \pi ^2 K^2 T^2}+1\right)}{K+1},
\ee
where $K=W\left(\frac{e^{-\frac{\phi }{8 \pi  T}}}{8 \pi  T}\right)$ and $W(x)$ is the Lambert function defined by the general formula $W(x)e^{W(x)}=x$.
As is depicted in Fig.\ref{XP}, in this case there is also a kind of criticality. However, the criticality is completely different from that of RN-AdS black hole. The plots of $\Xi-\phi$ for fixed temperature are something like the mirror symmetry of the plots of $T-S$ for fixed $\Xi$. However, when the value of $|\phi|$ is large, the isothermal curve with lower temperature has more larger value of $\Xi$. For infinite $|\phi|$, the curves all approach to $\Xi=1$.

%\section{Geometrothermodynamics of HL black hole}
%
%In this section, we employ the geometrothermodynamics (GTD)\cite{Quevedo} to characterize the thermodynamic stability and phase transition of the HL black hole. By incorporating the Legendre transformation into the geometric structure of the equilibrium space, a Legendre
%invariant thermodynamic metric can be constructed\cite{Quevedo1,Quevedo2,Quevedo3,Quevedo4}.
%We can introduce a five-dimensional
%phase space $\mathcal {T}$ with coordinates $\{\Psi,E^a,I^a\}=\{M,S,\Xi,T,\phi\}$, where $\Psi$ is the thermodynamic potential, $E^a$ and $I^a$ represent
% the extensive quantities and intensive quantities respectively. In the equilibrium space $\mathcal {E}\subset \mathcal {T}$, the induced metric
% can be constructed
% \be
% g=\left(E^c\d{\partial \Psi}{\partial E^c}\right)\left(\eta_{ab}\delta^{bc}\d{\partial^2 \Psi}{\partial E^c\partial E^d}dE^adE^d\right)
% \ee
%According to the first law, Eq.(\ref{1st}),
%\be
%g=(S \frac{\partial M}{\partial S}+\Xi \frac{\partial M}{\partial \Xi})(-\frac{\partial^2 M}{\partial S^2}dS^2+\frac{\partial^2 M}{\partial \Xi^2}d\Xi^2)
%\ee

\section{Concluding remarks}

In this paper we studied the thermodynamic stability of the HL black hole. We only consider the spherical horizon case and do not consider any matter fields. Without the detailed-balance condition, an extra parameter $\epsilon$ was introduced into the HL gravity. In the HL black hole solution, the parameter $\epsilon$ always occurs in the form $\epsilon^2$. Thus we let $\epsilon^2=\Xi$ for short.

By expanding the metric function of the HL black hole at large distance, we find that it has the similar structure as that of the RN-AdS black hole. So it is not surprise that the uncharged HL black hole has the $P-V$ criticality similar to that of RN-AdS black hole. However, in the HL black hole there are only two parameters $(M,~\Xi)$, while in the RN-AdS black hole there are three parameters $(M,~Q,~l)$. Therefore, in the peculiar $P-V$ criticality observed in \cite{Ma-PRD.2017}, the parameter $\Xi$ plays both the roles of the electric charge and the cosmological constant in the RN-AdS black hole. This feature leads to the occurrence of the critical curve, but not the critical point.

Further comparing the mass and temperature of the HL black hole with those of the RN-AdS black hole, we find they are also very similar. Thus, by analogy we treat the parameter $\Xi$ as a thermodynamic variable, such as $Q$ in the RN black hole. In this manner, we obtain the first law of thermodynamics for the HL black hole.  This analogy is not trivial. Because the entropies of the two black holes have very different forms, there must be some differences between their thermodynamic properties, and indeed it is. We find that the $T-S$ criticality for the two kinds of black holes have different behaviors. According to the $T-S$ curves of the HL black hole, we find a pseudo-VdW phase transition because $\beta=1/T$ will not tend to infinity as the entropy approaches zero. 

By calculating the heat capacity we find that the HL black hole is always thermodynamically stable when $\Xi<\Xi_c$. When $\Xi>\Xi_c$, the heat capacity will be locally thermodynamically stable in two disconnected regions. Further calculating the free energy, we find that the small HL black hole is more thermodynamically stable for lower temperature and the large HL black hole is more stable for higher temperature. We also studied the relations between $\Xi$ and its conjugated quantity $\phi$ for fixed temperature. It is shown that there is also a $\Xi-\phi$ criticality, but this criticality is very different from the $Q-\phi$ criticality for RN-AdS black hole.

\bigskip

%\section*{Acknowledgements}

\bibliographystyle{JHEP}

\end{CJK*}
\end{document}